\documentclass[dvips]{astavr}
\usepackage{supertabular,lscape,epsfig}
\usepackage{amssymb}
\usepackage{amsmath}

\SetPages{1}{1}
\SetVol{65}{2015}

\begin{document}

\begin{Titlepage}
\Title{33 Lib -- analog of gamma Equ }

\Author{V.D.~~B~y~c~h~k~o~v$^1$, L.V.~~B~y~c~h~k~o~v~a$^1$, J.~~M~a~d~e~j$^2$
          and~~G.~P.~T~o~p~i~l~s~k~a~y~a$^3$ }
{$^1$ Special Astrophysical Observatory of the Russian
        Academy of Sciences (SAO), Nizhnij Arkhyz, 369167 Russia \\
e-mail: (vbych,lbych)@sao.ru       \\
$^2$ Warsaw University Observatory, Al. Ujazdowskie 4, 00-478 Warszawa, Poland \\
e-mail: jm@astrouw.edu.pl          \\
$^3$  North Caucasian Federal University (NCFU), Pushkina 1, Stavropol, Russia \\
e-mail: gtop@mail.ru
}

\Received{Month Day, Year}
\end{Titlepage}

\Abstract{
In this paper we study periodic variability of the magnetic field in the
Ap star 33 Lib. We found that its most probable period equals 83.5 years.
There exist also possible shorter periods: 11.036 days, 7.649 days and 
4.690 days. Analysis of the magnetic behavior of 33 Lib allows us to conclude,
that the star shows the second longest period the slow rotator gamma Equ,
the latter star with the period of 97 years.
}
{Stars: chemically peculiar -- Stars: magnetic fields -- Stars: individual: 33 Lib }

\section{Introduction}

Magnetic field of chemically peculiar Ap star 33 Lib (HD 137949) was
discovered by H.W. Babcock on June 13, 1957 (Babcock 1958). Variability
of the longitudinal component $B_e$ of 33 Lib was first studied in 1971
(van den Heuvel 1971). Period of the magnetic variability was estimated
there, $P_{\rm mag} \approx 18.4$, based on a low number of measurements.
Furthermore, Wolff (1975) also found periods 23.26 days and 7.194 days.
from observations of the magnetic field. Most recent assessment of the 
magnetic period of 33 Lib was published by Romanyuk et al. (2014), who
determined $P_{\rm mag} = 7.0187$ days. Authors determined also the 
amplitude of $B_e$ variability equal 650 G with the average strength of 
the longitudinal field, $\overline{B_e} = 1528$ G.

Attempts to detect light and color variations of the star did not lead
to success (Wolff 1975; Deul \& van Genderen 1983). Researchers noted 
lack of brightness variations within accuracy of 0.01 magnitude. For 
the first time weak photometric variability with the period of 4.8511 
days was reported by Wraight et al. (2012). Data were obtained from by
the satellite STEREO (NASA), see Eyles et al. (2007). 

Kurtz (1982) discovered rapid photometric variability of 33 Lib with
the period $P = 8.2721$ minutes. This finding reinforced the analogy 
with another well known magnetic Ap star Gamma Equ. The latter object
exhibits strong global magnetic field, rapid photometric variability 
and long rotational period.

Ap stars provide a unique opportunity to assess unambiguously their
rotational periods of and angles between the axis of rotation and 
the line of sight. Thanks to that we discovered amazing objects -- 
"standing" or immobile stars with rotational periods equal to dozens 
or even hundreds of years old (Bychkov et al. 2006). Study of
long-period variable is always associated with the obvious problem: 
it is difficult to obtain a homogeneous series of measurements over
decades.

\section{Analysis of observational data }

Mathys et al. (1997) analysed variability of the longitudinal magnetic 
field of 33 Lib using measurements of $B_e$ obtained during 36 years 
and concluded, that the likely rotational period of the star equals
at least 75 years. Currently the available dataset contains more 
estimates of the magnetic field covering longer period of time - more 
than 50 years. Such an increase of time span allowed us to determine 
more realistic period of long term projected magnetic variations of
33 Lib, $P_{\rm mag}= 83.5 \pm 6.3$ years. The corresponding average
magnetic phase curve was presented in Figure 1. 

As can be seen, part of $B_e$ estimates significantly deviates from 
the average (sine wave) phase curve. We attempted to find a higher
order variations of the magnetic field. To do this, we have subtracted 
the average sine wave variation from the observed $B_e$ time series
and repeated spectral analysis of the prewhitened time series. Three 
new probable periods were found: 11.036, 7.649 and 4.690 days, which
are listed in the order of descending probability 2.87$\sigma$, 
2.80$\sigma$ and 2.65$\sigma$, respectively. Phase curves corresponding
to these periods are shown in Figs. 2--4. 

Estimates of the longitudinal magnetic field $B_e$ of 33 Lib were published
in Babcock (1958), van den Heuvel (1971), Wolff (1975), Hubrig et al. (2004),
Sachkov et al. (2011) and Romanyuk et al. (2014).

Sachkov et al. (2011) determined parameters and the chemical composition
of the atmosphere of 33 Lib on the basis of high resolution spectroscopy 
and calculations of model atmospheres and synthetic spectra. Analysis of
spectral energy distribution in the wavelength range 2000\AA -- 8000\AA
implied $T_{\rm eff}=  7550 \pm 100$ K and $\log g = 3.9 \pm 0.1$. These
parameters are parallax measured by satellite {\sc Hipparcos} implied
that the radius of 33 Lib equals $2.09 \pm 0.13 R_\odot $.

\begin{figure*}[h]
\includegraphics[angle=0, width=0.8\columnwidth]{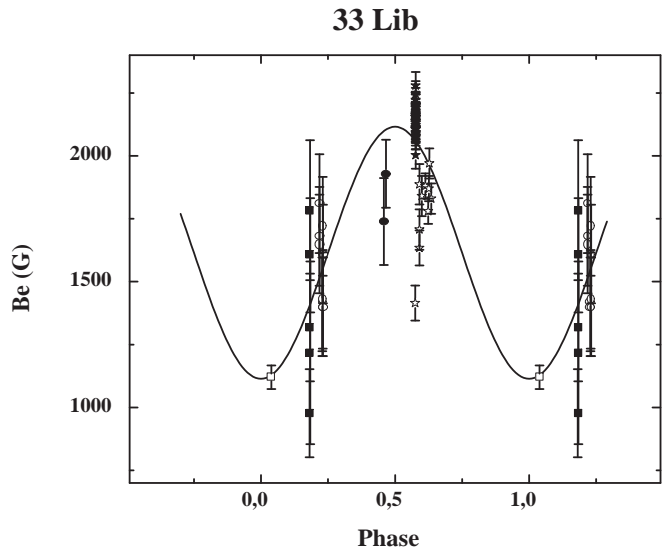}
\caption{ Magnetic phase curve computed for the period 83.5 years.}
\label{fig:1}
\end{figure*}

\begin{figure*}[htb]
\includegraphics[angle=0, width=0.8\columnwidth]{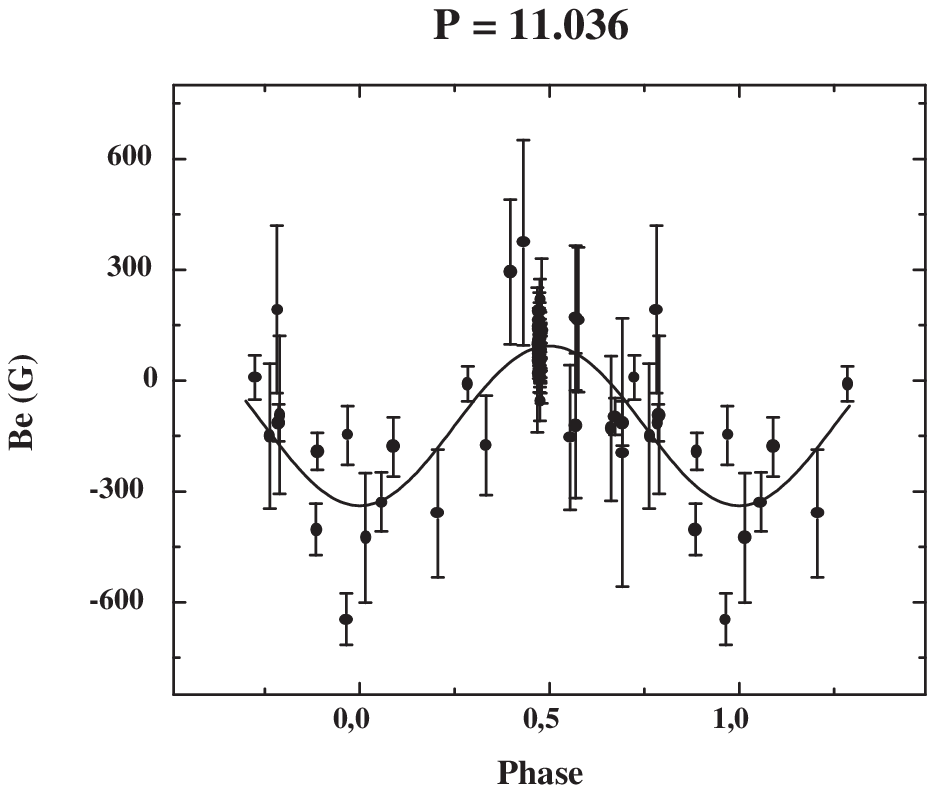}
\caption{Magnetic phase curve for 33 Lib found from deviations 
    for the period $P= 11.036$ days. }
\label{fig:2}
\end{figure*}

\begin{figure*}[h]
\includegraphics[angle=0, width=0.8\columnwidth]{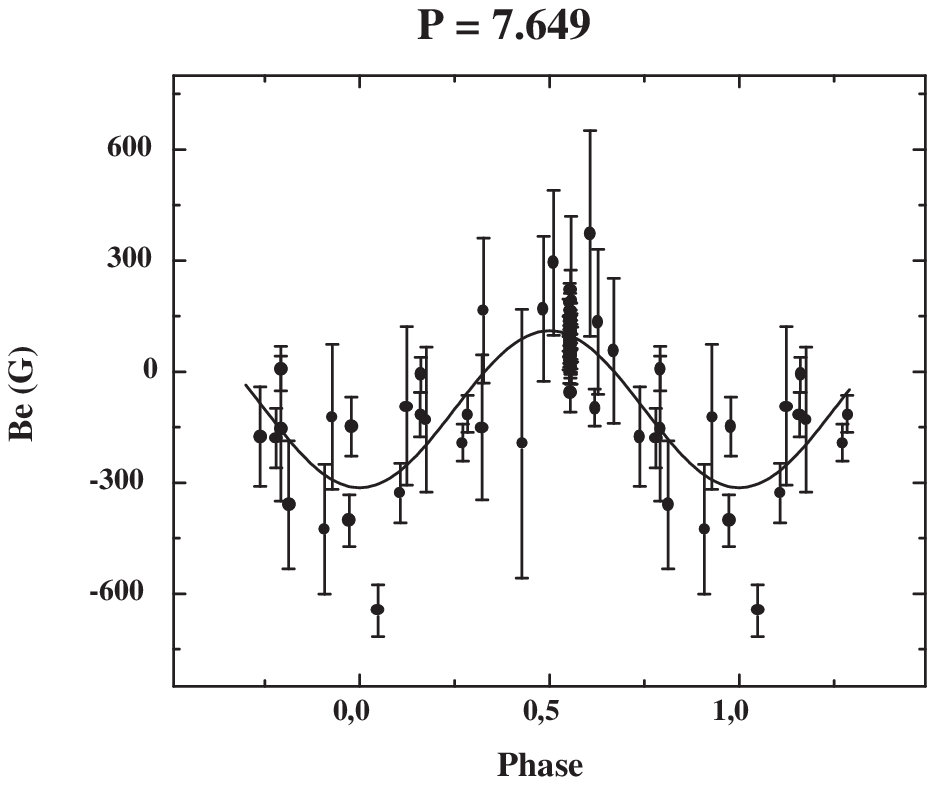}
\caption{Magnetic phase curve for 33 Lib found from deviations 
    for the period $P= 7.649$ days. }
\label{fig:3}
\end{figure*}

\begin{figure*}[htb]
\includegraphics[angle=0, width=0.8\columnwidth]{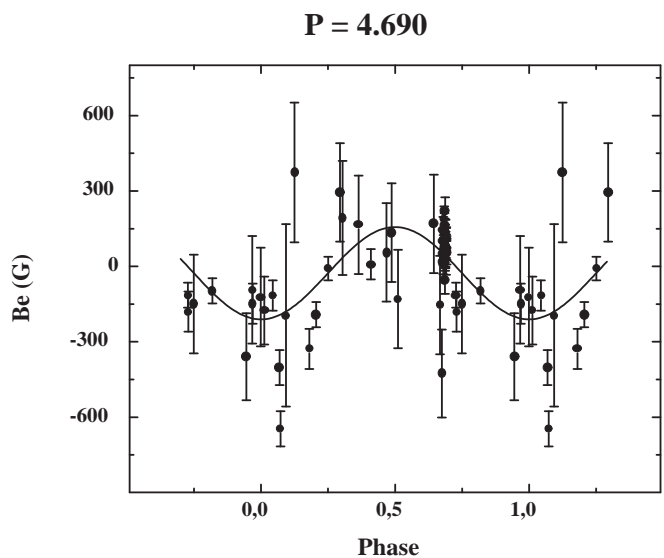}
\caption{Magnetic phase curve for 33 Lib found from deviations 
    for the period $P= 4.690$ days. }
\label{fig:4}
\end{figure*}

\section{Discussion}

Currently there are determined the following observational parameters 
and quantitative results of spectral analysis of 33 Lib:

\flushleft
\begin{enumerate}
\item Projection of the equatorial rotational velocity on the line of 
sight, $V \sin i = 0.6$ km/s (Ryabchikova, T.A., personal communication).
This value well correspond to our result, i.e. the period of the long 
term variations of the longitudinal magnetic field, $P_{\rm mag} = 83.5$
years. 
Relation between the rotational (or magnetic) period and the velocity of
axial rotation of the star on the equator is
\begin{equation}
V= 2\pi R/P = 1.3 \rm km/s \, . 
\end{equation}
Then, the resulting angle $i = 27^\circ $.
\item Significant deviations of the magnetic field $B_e$ points with 
respect to the average magnetic curve allowed us to assume, that there 
exist other possible  periods of variability of unknown nature. Those
periods equal 11.036, 7.649 and 4.690 days. It should be noted that the 
period of 7.649 reported here is close to the period of the longitudinal
magnetic field variation of 7.0187 days, found by Romanyuk et al. (2014).
Our suspected magnetic period of 4.690 days is close to the photometric 
period of 4.8511 days, independently found by Wraight et al. (2012). 
The most significant period of 11.036 days occurs first. The reasons 
for this short time variability are yet unknown.
\end{enumerate}

\section{Concluding remarks }

Actually there exists only 57 measurements of the longitudinal magnetic
field $B_e$ for 33 Lib, collected in time interval of about 50 years.
Note the uneven distribution of $B_e$ points in time, as well as the 
fact that the magnetic field estimates were obtained with various
methods and by different instruments. All those circumstances make an 
analysis of available data difficult. We hope that in the near future 
to increase the duration of available $B_e$ time series and the number 
of magnetic field observations. This would allow for more conclusive 
judgement regarding rotationaal period of the star 33 Lib.

\Acknow{
Authors are grateful for the support of the Polish National Science
Center grant number 2011/03/B/ST9/03281, the Russian grant "Leading Scientific
School" and Presidential grants MK-6686.2013.2 and MK-1699.2014.2.
Research by Bychkov V.D. was supported by the 
Russian Scientific Foundation grant N14-50-00043. }


\begin{references}

\refitem{Babcock, H.W.}{1958}{ApJS}{3}{141}
\refitem{Bychkov, V.D., Bychkova, L.V., Madej, J.}{2006}{MNRAS}{365}{585}
\refitem{Deul, E.R., van Genderen, A.M.}{1983}{A\&A}{118}{289}
\refitem{Eyles, C.J., et al.}{2009}{Solar Phys.}{254}{387}
\refitem{Hubrig, S., Kurtz, D.W., et al.}{2004}{A\&A}{415}{661}
\refitem{Kurtz, D.W.}{1982}{MNRAS}{200}{807}
\refitem{Mathys, G., et al.}{1997}{A\&AS}{123}{353}
\refitem{Romaniuk, I.I., et al.}{2014}{AstBu}{69}{427}
\refitem{Sachkov, M., et al.}{2011}{MNRAS}{416}{2669}
\refitem{van den Heuvel, E.P.}{1971}{A\&A}{461}{11}
\refitem{Wolff, S.C.}{1975}{ApJ}{202}{127}
\refitem{Wraight, K.T., et al.}{2012}{MNRAS}{420}{757}

\end{references}
\end{document}